\documentclass[apj]{emulateapj}
\newcommand{\be}{\begin{equation}}
\usepackage{threeparttable}
\usepackage{booktabs}
\newcommand{\ee}{\end{equation}}
\newcommand{\bea}{\begin{eqnarray}}
\newcommand{\eea}{\end{eqnarray}}

\usepackage{hyperref}
\usepackage{graphicx,times}
\usepackage{subfigure}
\usepackage{epstopdf}
\usepackage{amsmath,bm}
\usepackage{amssymb}
\usepackage{natbib}
\begin{document}

\title{Cosmic Ray Parallel and Perpendicular Transport in Turbulent Magnetic Fields}
\author{Siyao Xu\altaffilmark{1,2} and Huirong Yan\altaffilmark{1}}

\altaffiltext{1}{Kavli Institute of Astronomy and Astrophysics, Peking University, Beijing 100871, China; hryan@pku.edu.cn}
\altaffiltext{2}{Department of Astronomy, School of Physics, Peking University, Beijing 100871, China;  syxu@pku.edu.cn}

\begin{abstract}
A correct description of cosmic-ray (CR) diffusion in turbulent plasma is essential for many astrophysical and heliospheric problems. This paper aims to present physical diffusion behavior of CRs in actual turbulent magnetic fields, a model of which has been numerically tested. We perform test particle simulations in compressible magnetohydrodynamic turbulence. We obtain scattering and spatial diffusion coefficients by tracing particle trajectories. We find no resonance gap for pitch-angle scattering at 90$^\circ$. Our result confirms the dominance of mirror interaction with compressible modes for most pitch angles as revealed by the nonlinear theory. For cross-field transport, our results are consistent with normal diffusion predicted earlier for large scales. The diffusion behavior strongly depends on the Alfv$\acute{\rm e}$nic Mach number and the particle's parallel mean free path. We for the first time numerically derive the dependence of $M_A^4$ for perpendicular diffusion coefficient with respect to the mean magnetic field. We conclude that CR diffusion coefficients are spatially correlated to the local turbulence properties. On scales smaller than the injection scale, we find that CRs are superdiffusive. We emphasize the importance of our results in a wide range of astrophysical processes, including magnetic reconnection.
\end{abstract}


\section{INTRODUCTION}
Astrophysical plasma is generally turbulent due to the large spatial scales involved. Propagation of cosmic rays (CRs) in turbulent magnetic fields plays a key role in understanding many important issues both in space and astrophysics, e.g., solar modulation of CRs, CR acceleration, positron transport, CR anisotropy and diffuse $\gamma$-ray emission \citep[see][]{Jo_Pa69, YLS12}. 
However, CR diffusion in turbulent medium is still not fully understood. Current models on CR propagation are often developed by fitting observational data. The conventionally used assumption is that CR diffusion is isotropic and spatially homogeneous, but this too simplified assumption faces major problems in interpreting observations. In addition to the conventional problems, such as the ratio of the boron to carbon, mounting observation evidences challenge the traditional models of propagation. Examples include inconsistency between the EGRET data and locally measured spectra of CRs \citep{Strong04},
diffuse $\gamma$-ray excess in the inner Galaxy \citep{Ackermann12},
etc. All these observations imply that a spatially dependent diffusion may hold the key. Additional effects of turbulence on CR transport are discussed in some recent works 
\citep{Evo12, Toma12}.

CR diffusion depends on the turbulent magnetic fields adopted. Recent advances in turbulence studies necessitate corresponding revisions in CR transport theory. As revealed earlier, CR transport in tested models of turbulence is very different from earlier paradigm and is indeed inhomogeneous and can be anisotropic
(\citealt{YL02, YL04, YL08}, hereafter YL02, YL04 and YL08, respectively, see also a book by \citet{YL12book}). 
In this paper, we shall study numerically the transport of CRs in tested model of MHD turbulence 
(\citealt{GS95}, henceforth GS95; \citealt{LV99}; \citealt{CV00}; 
\citealt{CL03}; see review by \citet{Lazssrv09} and references therein).
We employ realistic turbulent magnetic fields, directly produced by three-dimensional MHD simulations, to provide a reliable description of the diffusion process of CRs. In particular, we shall use compressible MHD turbulence as our input for the following reasons. First of all, turbulence in nature is compressible with finite plasma $\beta\equiv P_{gas}/P_{mag}$. The magnetic pressure $P_{mag}$ cannot be neglected compared to gas pressure $P_{gas}$ for most of the medium that CRs propagate in \footnote{Otherwise without magnetic field the CRs' propagation would be ballistic, which is against what we know from observations.}. Second of all, the compressible modes, in particular fast magnetosonic modes, have been identified as the most important for CR scattering by both quasilinear theory (QLT, YL02, 04) and nonlinear theory (NLT, YL08). Indeed pseudo-Alfv$\acute{\rm e}$n modes (the incompressible limit of slow modes) can contribute through the mirror interactions. This process, however, does not function for particles with small pitch angles (YL08).

Perpendicular transport is another issue that we shall concentrate on in this paper. Many astrophysical environments including heliosphere and our Galaxy have well defined mean magnetic field. In spite of its fundamental importance, cross field transport remains an open question.  
A popular concept of CR cross field transport is subdiffusion 
\citep{Kota_Jok2000, Getmantsev, Mace2000, Qin2002, Webbcompound}.
But it fails to reproduce the diffusion process of solar energetic particles observed in the heliosphere 
\citep{Perri2009}. 
The solar energetic particle fluxes measured at different heliocentric distances indicate a faster diffusion process perpendicular to the solar magnetic field than subdiffusion 
\citep{Maclennan2001}.
Moreover, recent studies based on the tested GS95 model of turbulence have shown that subdiffussion does not apply and instead CR cross field transport is diffusive on large scales and superdiffusive on small scales 
(YL08; see review by \citet{Yan13}). 
On the contrary, superdiffusive behavior in the direction perpendicular to magnetic field, with displacement squared proportional to the third power of the distance along magnetic field, follows from the GS95 theory (\citealt{LV99}; \citealt{Narayan_Medv}; \citealt{LVC04}; \citealt{Maron04}; \citealt{Lazarian06}; YL08). This superdiffusion is important for, e.g., particle acceleration 
\citep{LY13}.

In this work, we will focus on investigating the diffusion process of CRs based on the tested model of turbulence. The structure of the paper is as follows. In Section 2, we describe the turbulent magnetic fields we use. In Section 3, we perform test particle simulations in the generated magnetic fields. We investigate particle scattering and parallel diffusion processes in Section 4. In Section 5\&6,  we present the results on particle perpendicular transport, followed by discussions and summary in Section 7\&8.

\section{Generation of turbulent magnetic fields}
We use the 
\citet{CL02_PRL}
code to generate isothermal compressible MHD turbulence at $512^{3}$ resolution. We drive the turbulence solenoidally in Fourier space with the energy injection scale $L$ equal to 0.4 cube size. The turbulence evolves on a Cartesian grid with mean magnetic field along the x-direction. We set initial density and velocity fields to unity, and adopt the same initial gas pressure value for all our simulations. The total magnetic field is a sum of a uniform background component and a fluctuating component, $\bm B=\bm B_{\rm ext}+\bm b$.  Initially we have $\bm b=0$, and $B_{\rm ext}$ is the only controlling parameter in our MHD simulations. By varying the external magnetic field values $\bm B_{\rm ext}$, we derive a data set of MHD turbulence with different Alfv$\acute{\rm e}$nic Mach numbers. The Alfv$\acute{\rm e}$nic Mach number is 
\begin{equation}
                M_A\equiv\left \langle |\bm v|/v_A \right \rangle,    
\end{equation}
where $\bm v$ is the local velocity, $v_A=|\bm B|/\sqrt{\rho}$ is the Alfv$\acute{\rm e}$nic velocity, $\bm B$ is the local magnetic field, and $\rho$ is density. Here $\left \langle...\right \rangle$ means a spatially averaged value over all grid points.

\begin{figure*}[htbp]   
    \subfigure[]{
     \includegraphics[width=6cm]{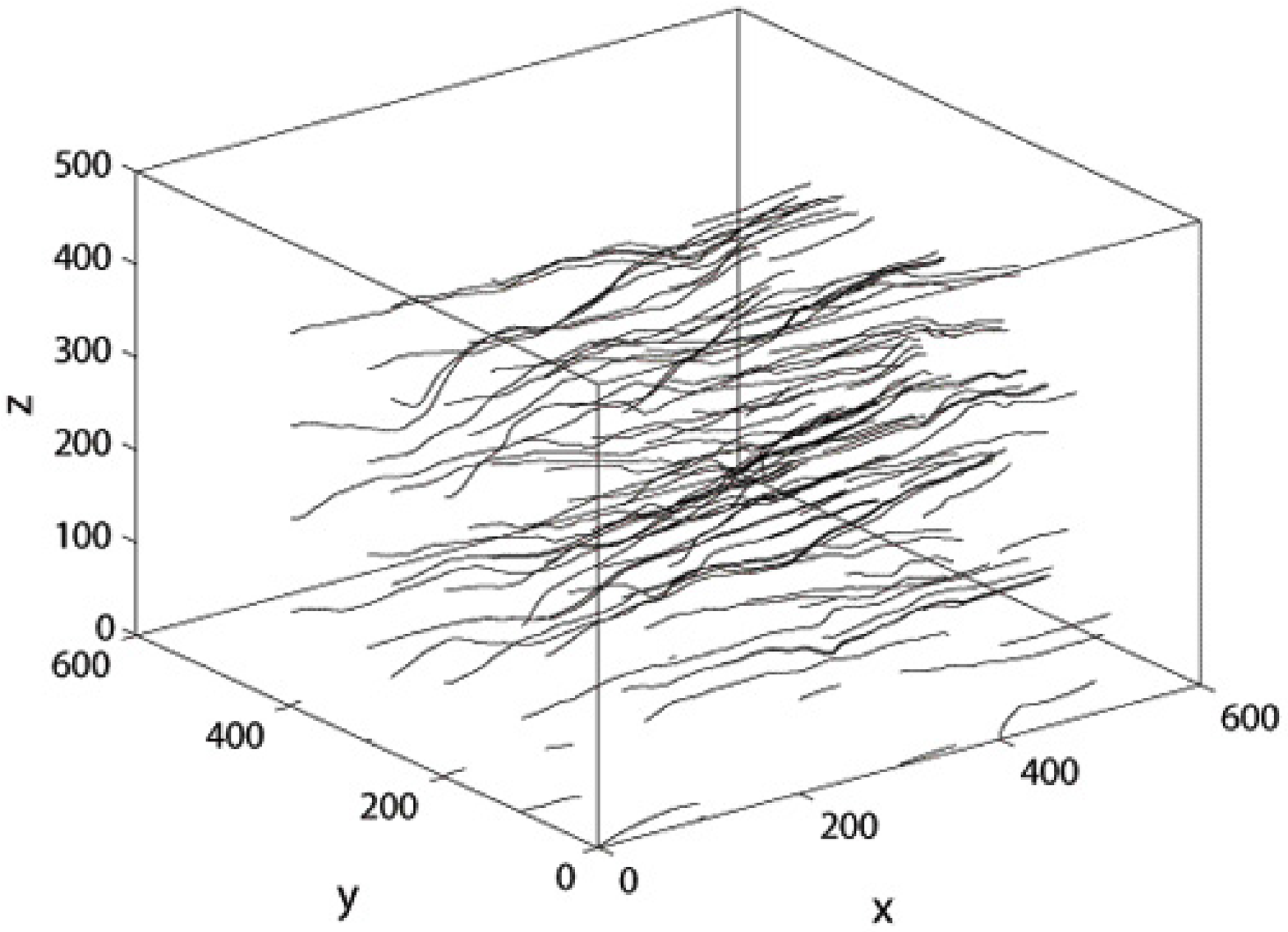}\label{figmf3}}
    \subfigure[]{
     \includegraphics[width=6cm]{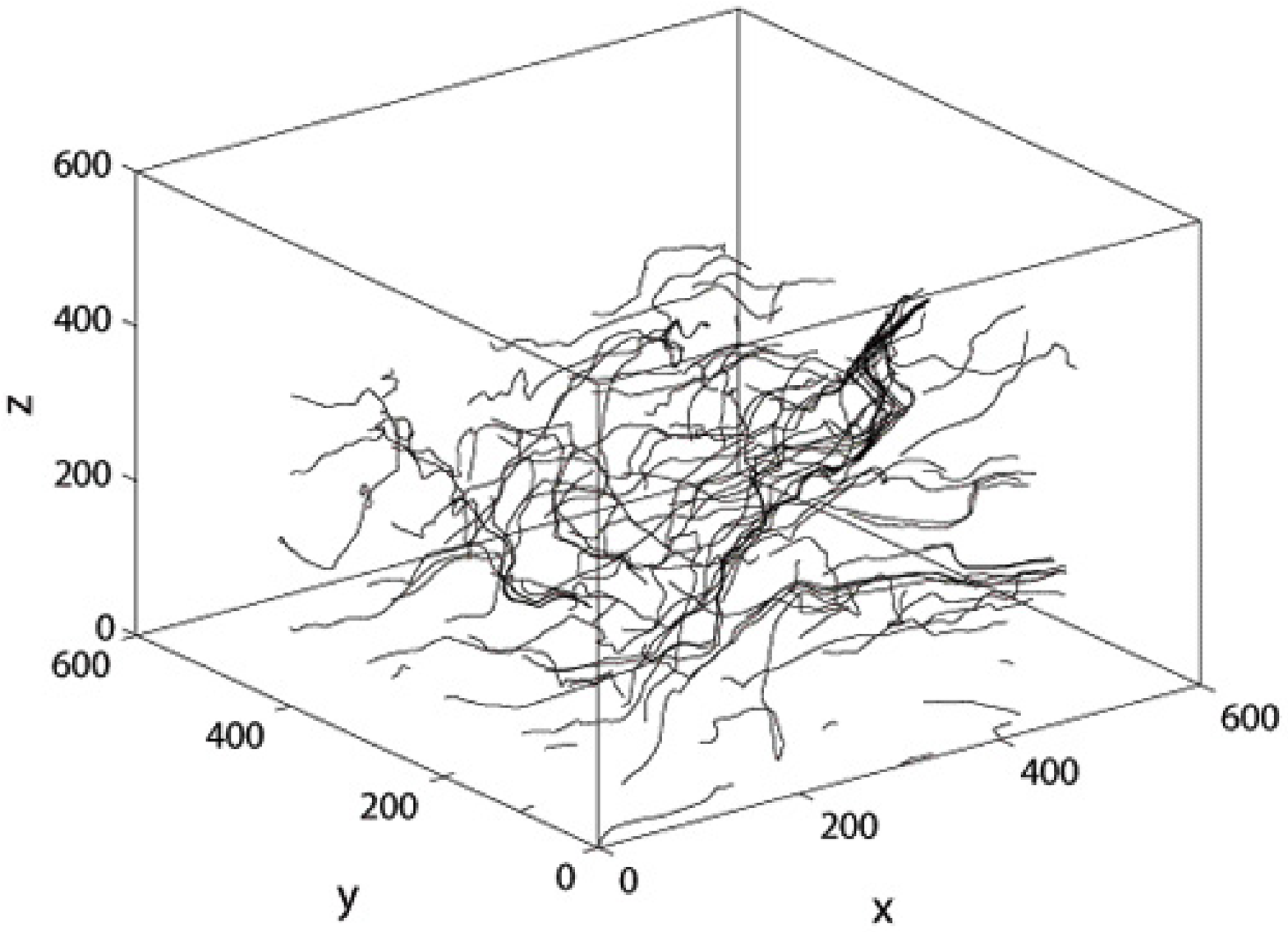}\label{figmf6}}
    \subfigure[]{
     \includegraphics[width=6cm]{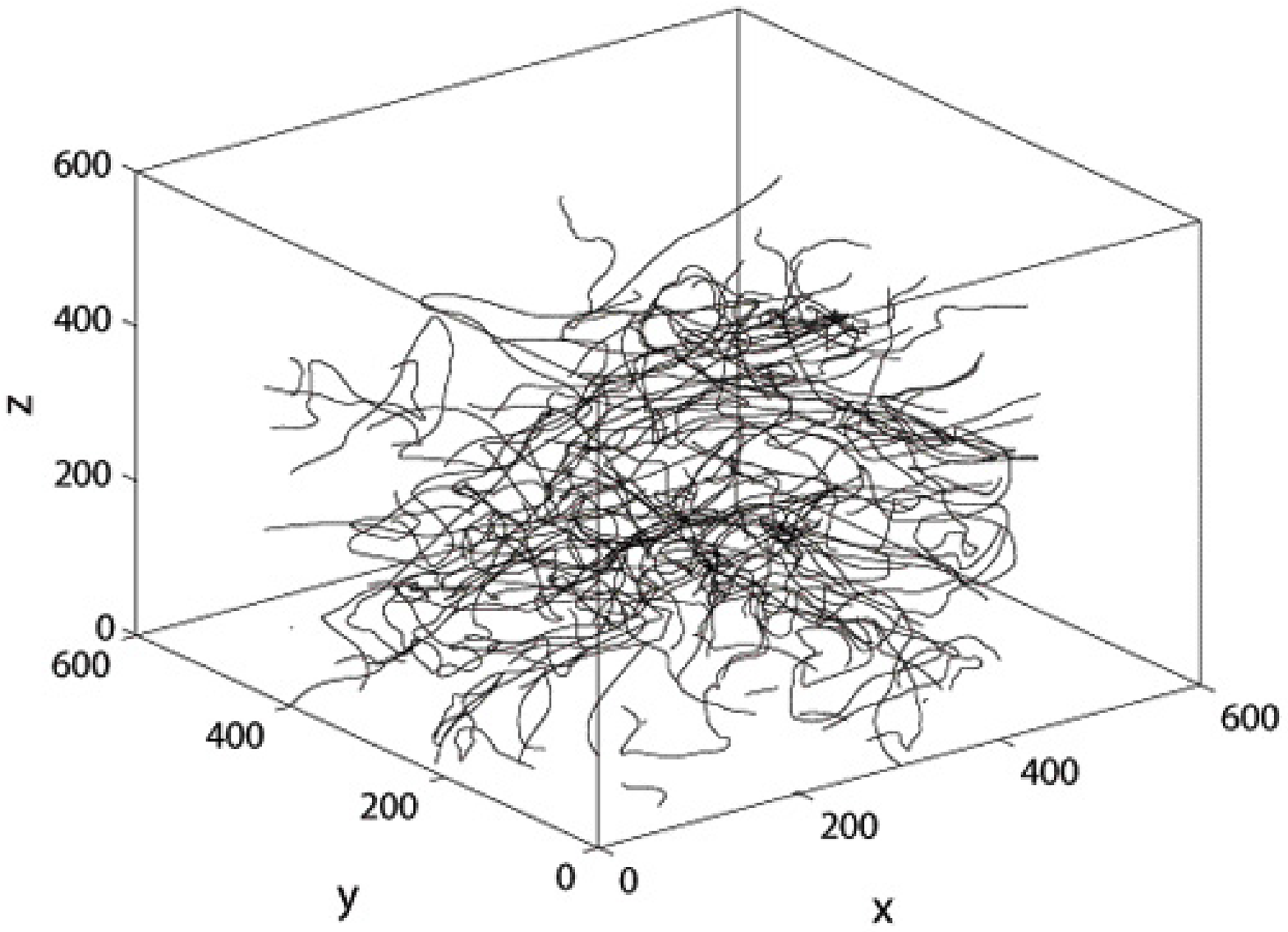}\label{figmf15}}
   \caption{\small Examples of generated magnetic fields from MHD simulations with different Alfv$\acute{\rm e}$nic Mach numbers, (a) $M_A=0.30$, (b) $M_A=0.73$ and (c) $M_A=1.5$. The lines display the magnetic field stream lines.}
\end{figure*}

$M_A$ describes the perturbation strength of the turbulence with respect to the mean field, and is the single parameter that characterizes the magnetic fields we use. Fig~\ref{figmf3}, Fig~\ref{figmf6}, and Fig~\ref{figmf15} display examples of resulting magnetic fields with the same input parameters except for different $\bm B_{\rm ext}$ values. These magnetic fields clearly have different structures and $M_A$ values. We divide our data into sub-Alfv$\acute{\rm e}$nic ($M_A<1$) and super-Alfv$\acute{\rm e}$nic ($M_A>1$) turbulence for the following test particle simulations.  Table~\ref{tab1} lists the $M_A$ values for the magnetic fields we use in this work.
\begin{table}
\centering
\begin{threeparttable}
\caption[]{$M_A$ values of the magnetic fields produced by MHD simulations}\label{tab1} \setlength\tabcolsep{2.3pt}
  \begin{tabular}{c|cccccccccc|c}
      \toprule
       & \multicolumn{10}{|c|}{Sub-Alfv$\acute{\rm e}$nic} & Super-Alfv$\acute{\rm e}$nic \\
      \midrule
        $M_A$ & 0.19 & 0.27 & 0.30 & 0.41 & 0.47 & 0.49 & 0.54 & 0.61 & 0.68 &  0.73 & 1.5 \\
       \bottomrule
    \end{tabular}
 \end{threeparttable}
\end{table}

\section{Test Particle Simulations}
After the MHD turbulence is fully developed, we use snapshots of turbulence separated by $\gtrsim$ the turnover time of the largest eddy as different magnetic field realizations. We trace the trajectories of CRs in the test particle simulations. Since the relativistic particles have speed much larger than the Alfv$\acute{\rm e}$n speed, the magnetic field can be treated as stationary and the electric field in the turbulent plasma can be safely neglected for the study of transport of CRs. We use the Bulirsh-Stoer method 
\citep{Press86}
to trace the trajectories of test particles. The algorithm uses an adaptive time-step method and the particle energy is conserved to a high degree during the simulation. 

In each time step, the magnetic fields defined on grid points are interpolated to the position of a test particle using a cubic spline routine. Given the local magnetic field $\bm B$, the trajectory can be computed by integrating the Lorentz force on each particle,
\begin{equation}
         \frac{d\bm u}{dt} = \frac{q}{mc}\bm u \times \bm B,      
\end{equation}
where $\bm u$ is the particle's velocity and the remaining symbols have their standard meanings.
We also use periodic box boundary conditions to keep the number of test particles unchanged.

In each simulation, we release 1000 test particles with random initial positions and pitch angles through the simulation cube. The particle energy is represented by its Larmor radius, expressed as 
\begin{equation}
         r_L=\frac{u}{\Omega}.
\end{equation}
Here $\Omega$ is the frequency of a particle's gyromotion,
\begin{equation}
         \Omega=\frac{eB}{\gamma mc},
\end{equation}
where $\gamma$ is the particle's gamma-factor.

To examine the effects of sample size on statistics, we perform test particle simulations with different number of particles. We show the results for perpendicular diffusion coefficient $D_\perp $ (in units of $\Omega^{-1}$) in Table~\ref{tab2} as an example. We will discuss the measurement for $D_\perp $ in detail below. For magnetic fields with different $M_A$, $D_\perp $ will always become stable when the sample size reaches $\sim1000$. Our tests show that the statistics will not depend on the sample size when test particle numbers are equal (or larger than) 1000. Thus we use 1000 as our sample size in the following test particle simulations. 

\begin{table}
\centering
\begin{threeparttable}
\caption[]{$D_\perp/\Omega$ derived from test particle simulations with different sample sizes. }\label{tab2} \setlength\tabcolsep{2.3pt}
  \begin{tabular}{cccc}
      \toprule
        Sample size & $M_A=0.30$ & $M_A=0.54$ & $M_A=0.73$  \\
      \midrule
        $100$              & $1.25e-6$        &  $ 1.80e-5 $        &    $   1.22e-4 $  \\
        $500$             & $1.56e-6  $      &   $1.70e-5   $      &     $  1.31e-4  $ \\
        $1000$            & $1.59e-6 $      &   $ 1.76e-5  $      &     $ 1.31e-4  $ \\
        $1500$           & $1.62e-6  $    &    $1.77e-5   $     &     $  1.28e-4  $ \\
        $2000$            &  $1.61e-6 $    &    $1.76e-5  $      &     $  1.29e-4  $ \\
       \bottomrule
    \end{tabular}
 \end{threeparttable}
\end{table}

\begin{figure*}[htbp]
   \centering
   \subfigure[]{
   \includegraphics[width=7.5cm]{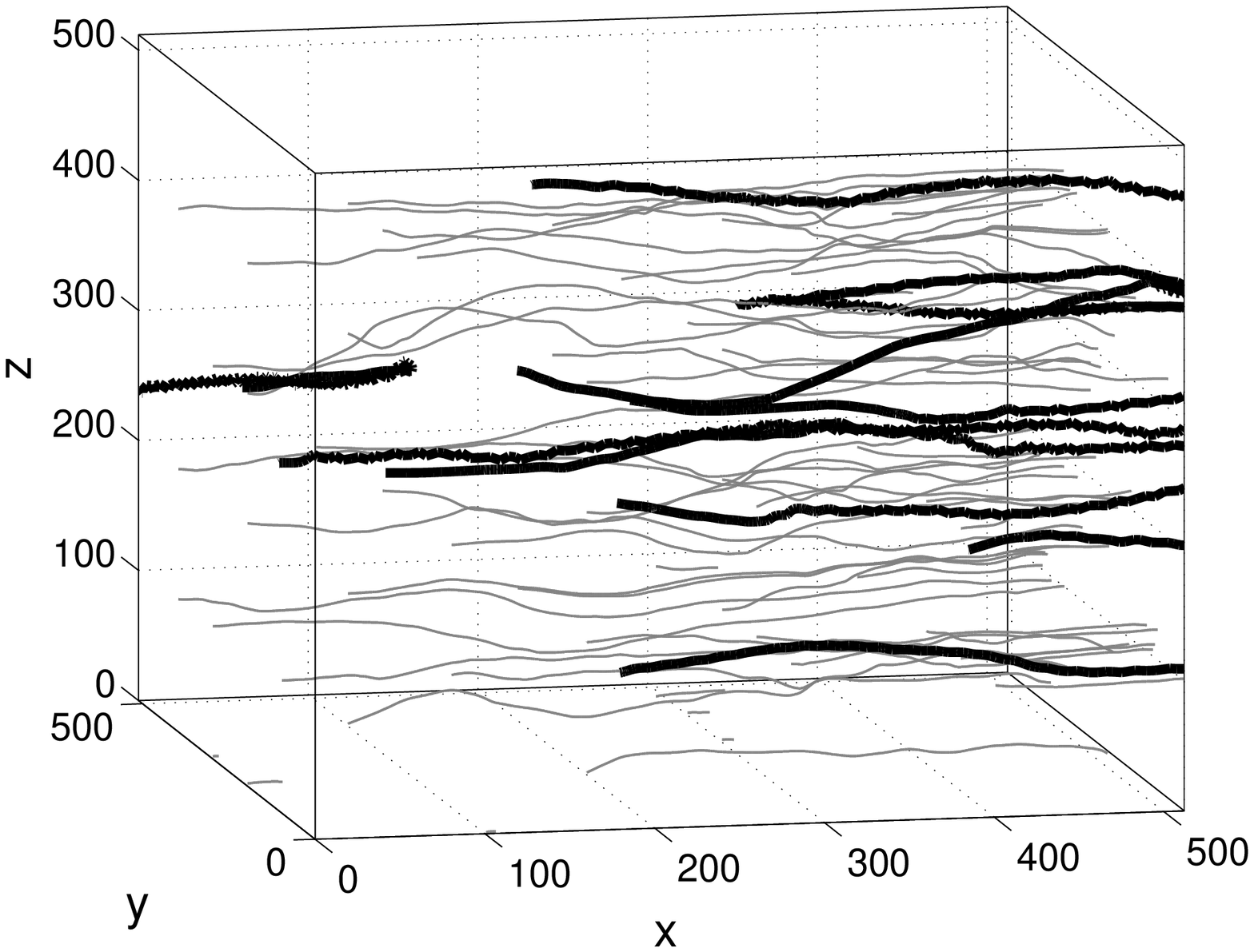}\label{fig2a}}
   \subfigure[]{
   \includegraphics[width=8cm]{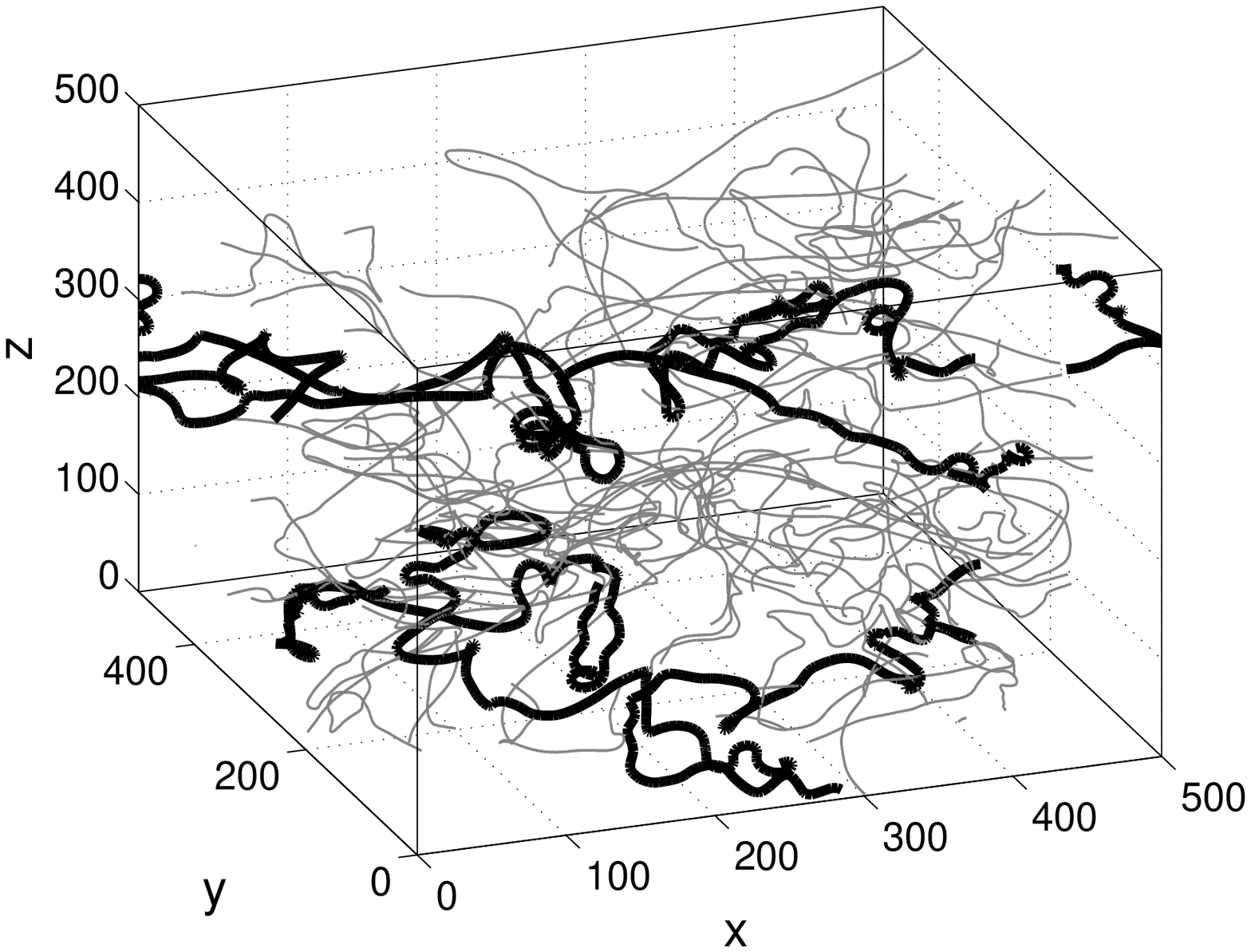}\label{fig2b}}
   \caption{\small Particle trajectories (thick black lines) in (a) sub-Alfv$\acute{\rm e}$nic turbulence with $M_A=0.30$ and (b) super-Alfv$\acute{\rm e}$nic turbulence with $M_A=1.5$. The thin gray lines display the magnetic field stream lines.}
\end{figure*}

Fig.~\ref{fig2a} and \ref{fig2b} show sample particle trajectories in three-dimensional turbulent magnetic fields with different $M_A$. Obviously, particle diffusion strongly depends on the properties of the turbulence.

\begin{figure*}[htbp]
   \centering
   \subfigure[]{
   \includegraphics[width=8cm]{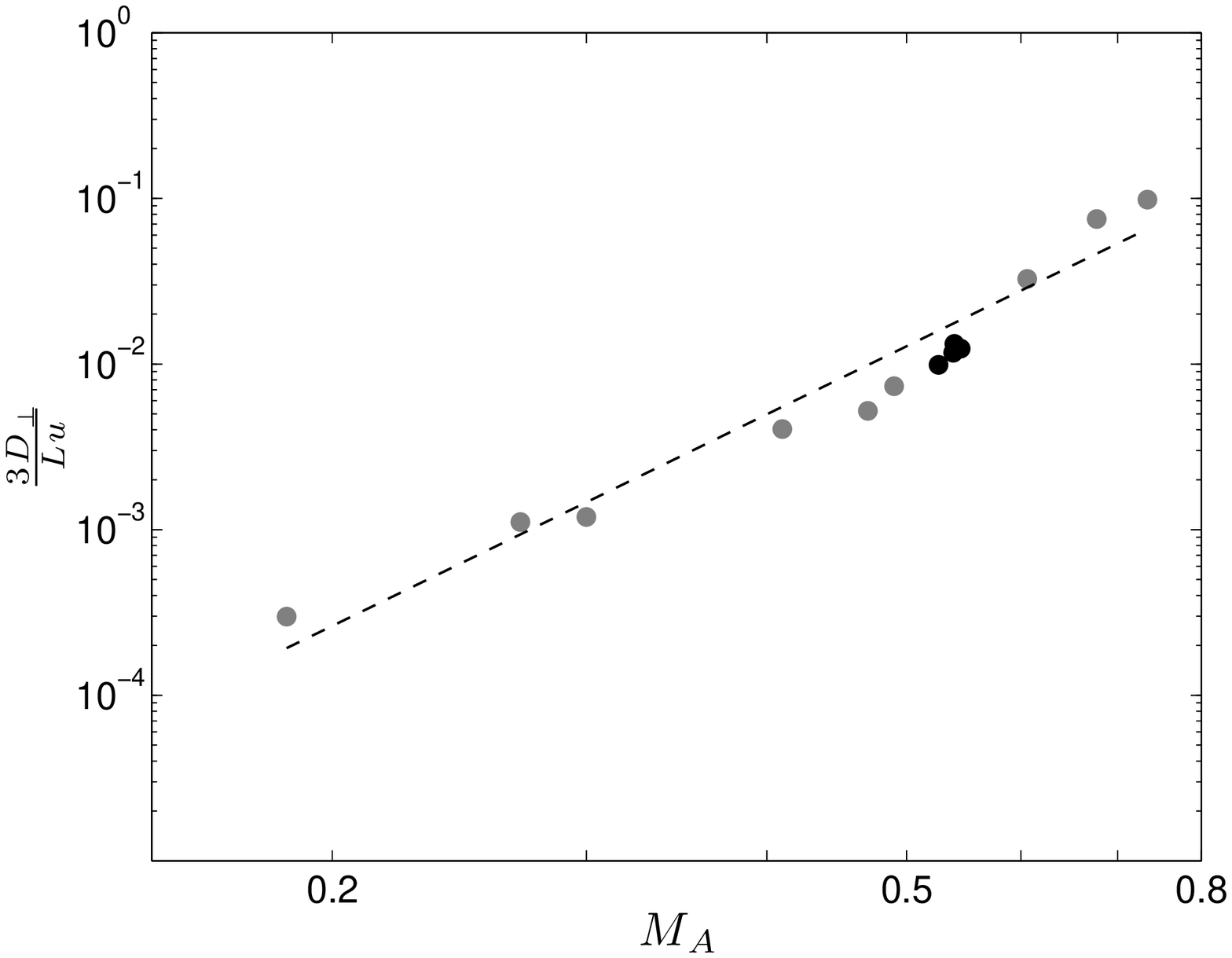}\label{figmagr1}}
   \subfigure[]{
   \includegraphics[width=8cm]{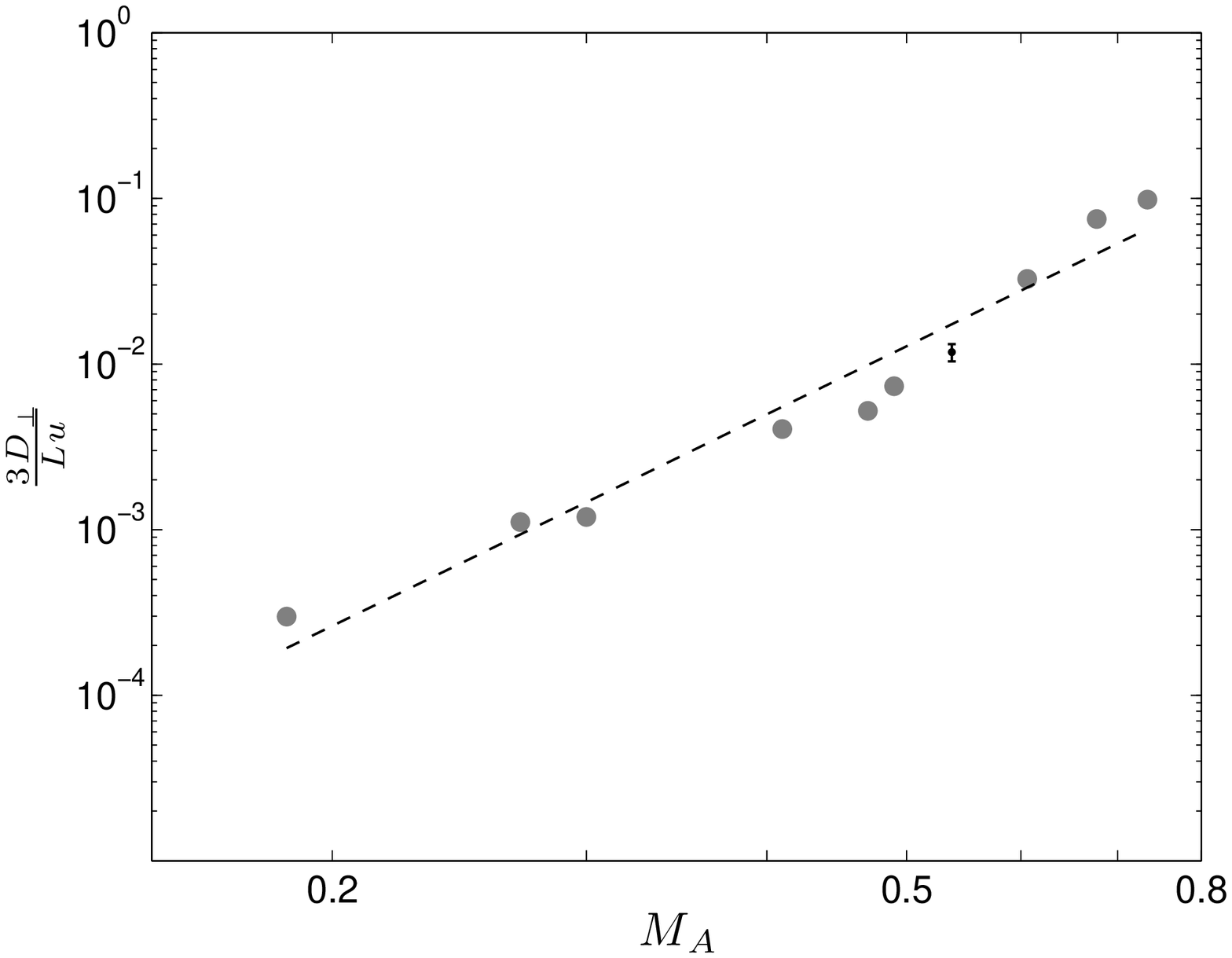}\label{figmagr2}}
   \caption{\small $\frac{3D_\perp}{Lu}$ as a function of $M_A$.The dashed line shows the best fit to the numerical results (filled circles). (a) Black circles are results for different snapshots of one magnetic field data set, and gray circles are results using a single snapshot of magnetic field. (b) Same as (a), except that the small-size dot at $M_A=0.54$ is the average value of the four data points (black circles in (a)). The error bar indicates the standard deviation of the four values.}
\end{figure*}

To examine the variations of numerical results arising from different magnetic field realizations, we perform test particle simulations in its different magnetic field realizations with a constant $M_A$. Fig.~\ref{figmagr1} shows the $D_\perp$ results derived from four snapshots of magnetic field with an average $M_A$ value equal to $0.54$ (black circles)\footnote{Note that it is not possible to generate turbulence data with exactly the same $M_A$ because of statistical fluctuations}, along with the results from other magnetic field data using a single snapshot (grey circles).  The black circles are overlapped due to the marginal difference in $D_\perp$ values. The best fit to the data (dashed line) has a slope of $4.11\pm0.66$ with a 95\% confidence level. In Fig.~\ref{figmagr2} , we show $D_\perp$ averaged from the four values using different magnetic field realizations and the error bar calculated from the standard deviation. Since the error bar has a height comparative to the symbol size, we use a small-sized black dot to exhibit the mean $D_\perp$ value. Other symbols are the same as those in Fig.~\ref{figmagr1}. The slope of the best fit changes slightly, to $4.21\pm0.75$ with a 95\% confidence level. We can clearly see that different realizations of magnetic fields only induce marginal difference to the results. So we can safely neglect this effect in our statistical analysis.

\section{Pitch-angle scattering and parallel diffusion}
\begin{figure}[htbp]
   \centering 
      \includegraphics[width=8cm]{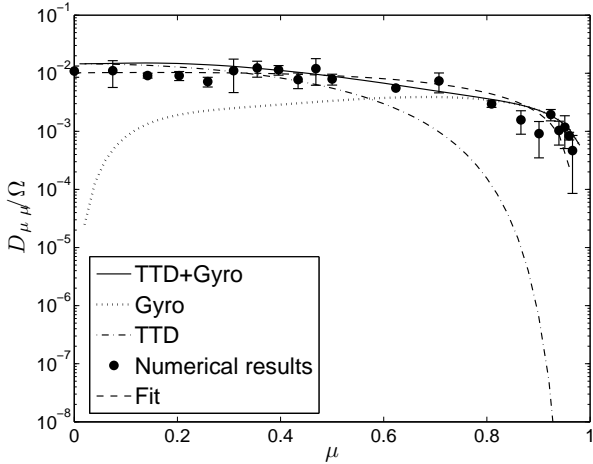}
   \caption{\small  Pitch angle diffusion coefficients (filled circles) measured for different initial pitch angles. The numerical results are fitted with a smooth curve (dashed line). Dash-dotted line and dotted line refer to the contribution from TTD and gyroresonance respectively (from YL08). The solid line represents the sum of them. }
   \label{fig4}
\end{figure}

We perform the scattering experiments using an ensemble of particles with a specific pitch angle at the starting point. During the scattering process, the pitch angle, i.e. the angle between the particle's velocity vector and the local magnetic field direction, changes with time. We trace the change of the pitch-angle cosine $(\mu-\mu_0)$ in a short time interval to keep the deviation of $\mu$ small 
\citep[see][]{BYL2011}, 
and obtain pitch-angle diffusion coefficient $D_{\mu\mu}$ by using the definition,
\begin{equation}
        D_{\mu\mu} =\frac{\left \langle (\mu-\mu_0)^2 \right \rangle}{2t},
\end{equation}
averaged over the ensemble of particles. Here $\mu_0$ and $\mu$ are the initial and final pitch-angle cosine respectively, and $t$ is the integration time. Fig.\ref{fig4} displays the measured $D_{\mu\mu}$ for particles with the same energy, $r_L=0.03$ cube size and different $\mu_0$ in the turbulence with $M_A=0.54$. Error bars in Fig.\ref{fig4} and the following figures are associated with variance of the Monte Carlo simulations. The fitting $D_{\mu\mu}$ curve smoothly extends from $\mu_0=0$ to $\mu_0\sim1$. Particles with a wide range of pitch angles, including 90$^\circ$, are scattered due to the resonance broadening, in contrast to the quasilinear theory results.
In quasilinear theory, mirror resonance has a sharp peak at large pitch angles close to 90$^\circ$, but is zero at 90$^\circ$ because of the discrete resonant Landau resonance condition $k_\parallel u_\parallel=kv_A$, where $k_\parallel$ is the component of the wavevector $\bm k$ parallel to the mean magnetic field,  and $u_\parallel$ is the parallel velocity component of a particle. In the mean time, gyroresonance also does not function at 90$^\circ$ according to its resonance condition. In nonlinear theory, nonetheless, the small gap around 90$^\circ$ disappears because of the resonance broadening. Fig.~\ref{fig4} also displays the pitch-angle scattering arising from QLT calculation of gyroresonance (Gyro)\footnote{The study in YL08 shows that the difference between QLT and NLT is marginal for gyroresonance, which operates only with small scale fluctuations unlike TTD. and transient time damping (TTD)  interaction calculated with NLT separately, and the total contribution of them. These results are from the analytical work predicted in YL08.} It is clear from Fig.~\ref{fig4} that our result agrees well with the prediction of the nonlinear theory in YL08. Their analytical calculations show that mirror interaction dominates for large pitch angles till 90$^\circ$.

The pitch-angle scattering determines the diffusion of CRs parallel to the magnetic field. By substituting $D_{\mu\mu}$ we measured into the equation
\citep{Earl:1974}
\begin{equation}
         \frac{\lambda_\parallel}{L}=\frac{3}{4}\int_{0}^{1}d\mu_0\frac{u(1-\mu_0^2)^2}{D_{\mu\mu}L},
\label{mfp}
\end{equation}
where $u$ is particles' velocity, we can obtain the corresponding parallel mean free path of particles. For instance, the corresponding mean free path is $\lambda_\parallel\approx1.3$ in cube size units for the case considered in Fig.~\ref{fig4}.

\begin{figure}[htbp]
   \centering 
      \includegraphics[width=8cm]{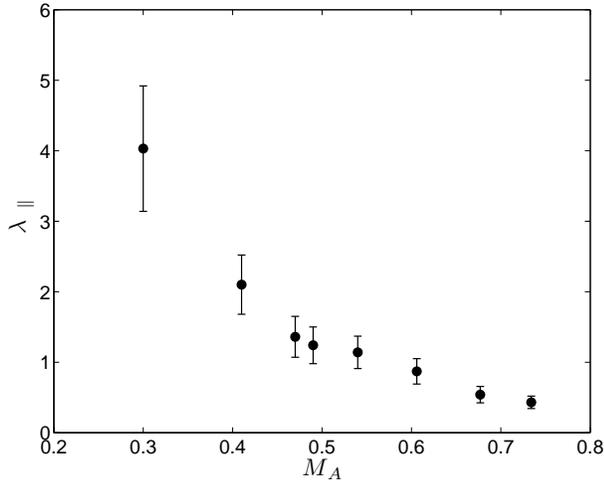}
     \caption{\small  $\lambda_\parallel$ (in cube size units)  for particles with $r_L=0.01$ cube size as a function of $M_A$. $\lambda_\parallel$ values are deduced from $D_\parallel$.}
      \label{fig5}
\end{figure}

To measure the parallel diffusion coefficient, we trace the particles over a long distance until we find that they enter the normal diffusion regime, i.e.,
\begin{equation}
        \left \langle (\tilde{x}-\tilde{x_0})^2 \right \rangle \propto t.   
\end{equation}
$(\tilde{x}-\tilde{x_0})$ is the distance measured parallel to the local magnetic field, and then we take the averaged square distance over all particles. 
The diffusion coefficient is calculated following the definition
\citep{Giacalone_Jok1999},
\begin{equation}
         D_\parallel =\frac{\left \langle (\tilde{x}-\tilde{x_0})^2 \right \rangle}{2t}.
\end{equation}
Given the parallel diffusion coefficient, we compute the parallel mean free path $\lambda_\parallel$ of particles directly from $D_\parallel$ using the relation
\begin{equation}
         \lambda_\parallel=\frac{3D_\parallel}{u}.
\end{equation}
Fig.~\ref{fig5} displays the resulting $\lambda_\parallel$ for particles with $r_L=0.01$ cube size vs. $M_A$. We find $\lambda_\parallel$ decreases with $M_A$, showing the increased $M_A$ leads to an enhanced efficiency in particle scattering. 

\section{CR perpendicular transport on large scales}
Since the properties of CR perpendicular diffusion strongly depend on the scale, namely, whether it is larger or smaller than the correlation length of turbulence $L$,  we consider space diffusion separately on large and small scales.

On large scales, similar to the parallel diffusion we described above, we observe that
\begin{equation}
        \left \langle (y-y_0)^2 \right \rangle \propto t,
\end{equation}
where $(y-y_0)$ represents the perpendicular distance, and 
\begin{equation}
         D_\perp=\frac{\left \langle (y-y_0)^2 \right \rangle}{2t}.
\end{equation}
Note that the perpendicular diffusion coefficient $D_\perp$ is calculated across the average magnetic field in the global frame of reference. 

For the super-Alfv$\acute{\rm e}$nic turbulence, it is straightforward to see that the transport is isotropic with a uniform diffusion coefficient since there is no mean magnetic field. Thus we focus on the sub-Alfv$\acute{\rm e}$nic turbulence ($M_A<1$). 

\begin{figure}[htbp]
      \includegraphics[width=8cm]{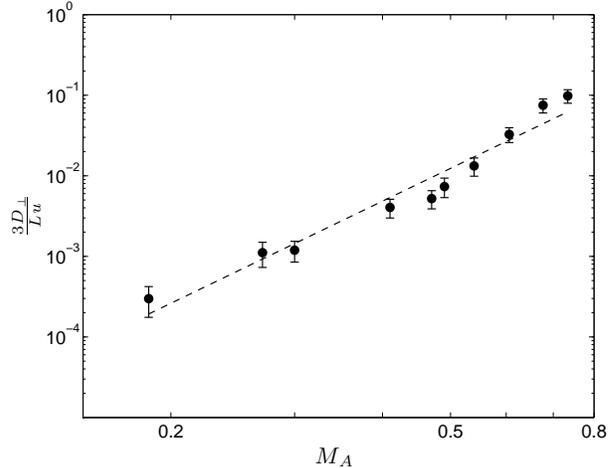}
   \caption{\small  $\frac{3D_\perp}{Lu}$ as a function of $M_A$.The dashed line shows the best fit to the numerical results (filled circles).}
   \label{fig6}
\end{figure}

In the sub-Alfv$\acute{\rm e}$nic turbulence, the mean free paths of the test particles are large because of low scattering rate (see Fig.~\ref{fig5}). Due to the limited inertial range of the current MHD simulations, $\lambda_\parallel$ is larger than the injection scale $L$ even for the particles of lowest  energies attainable. Thus we consider the case of $\lambda_\parallel>L$,  for instance, the cases of ultra high energy CRs and the transport of high energy Galactic CRs in small scale interplanetary turbulence. We measure the perpendicular diffusion coefficients of particles propagating in sub-Alfv$\acute{\rm e}$nic turbulence with different $M_A$. Fig.~\ref{fig6} presents $D_\perp$ of particles with $r_L=0.01 $ cube size as a function of $M_A$.  The results can be fitted by a line with a slope of $\approx 4.21$, indicating
\begin{equation}
         D_\perp \propto M_A^{4.21\pm0.75},
\end{equation}
with a 95\% confidence level. This relation is consistent with equation (26) in YL08, and confirms the dependence of $M_A^4$ instead of the $M_A^2$ scaling in, e.g., 
\citet{Jokipii1966}. This is exactly due to the anisotropy of the Alfv$\acute{\rm e}$nic turbulence. In the case of sub-Alfv$\acute{\rm e}$nic turbulence, the eddies become elongated along the magnetic field from the injection scale of the turbulence 
(\citealt{Lazarian06}; YL08). 
The result indicates that CR perpendicular diffusion depends strongly on $M_A$ of the turbulence, especially in magnetically dominated environments, e.g., the solar corona.

\section{Perpendicular Transport on Small Scales}

\begin{figure}[htbp] 
   \centering 
      \includegraphics[width=8cm]{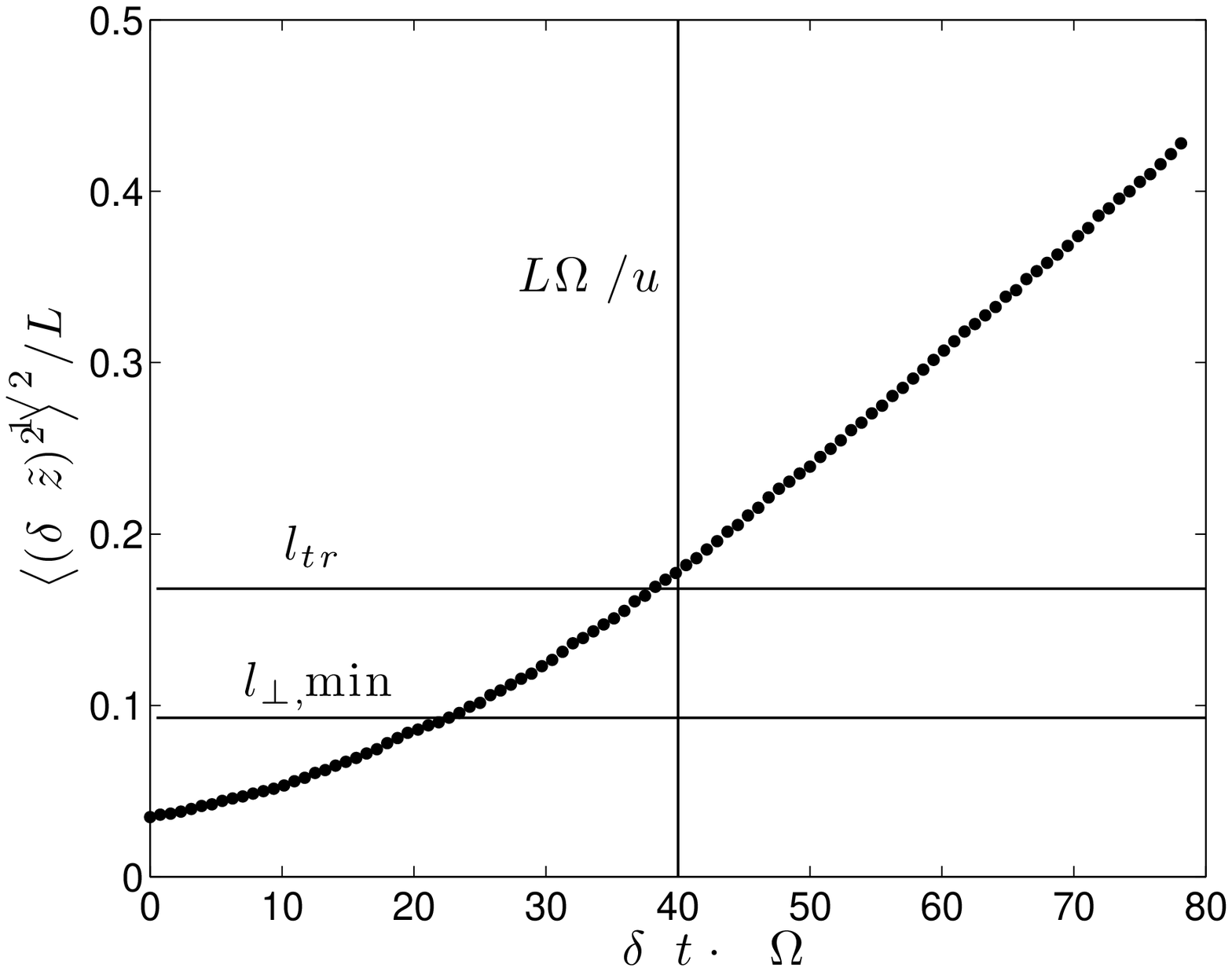}
   \caption{\small $\left \langle (\delta \tilde{z})^2 \right \rangle^{1/2}/L$ vs. $\delta t \cdot \Omega$ for $M_A=0.41$. The vertical line denotes the time for particles to travel $L$ along the direction of magnetic field. }
   \label{figev}
\end{figure}

We next consider the perpendicular transport specifically at scales smaller than the injection scale $L$. We simultaneously inject 40 beams of test particles randomly in the simulation cube. There are 20 particles in each beam. The spatial separations between their initial positions are equal to several grids, and their initial pitch angles are equal to $0^{\circ}$. All the particles have the same energy with $r_L=0.01$ cube size. At each timestep of the particle simulation, we measure the distances between particle trajectories as $\delta \tilde{z}$. And we take the rms of this value $\left \langle (\delta \tilde{z})^2 \right \rangle$ as the perpendicular displacement in the following discussions. We use the same method described in \citet{LVC04}, except that here we deal with particle trajectories instead of magnetic field lines. \footnote{The relation between the concept of magnetic field lines and the particle trajectories that trace magnetic field lines is discussed in  detail in \citet{Eyink2011}.} Fig.~\ref{figev} shows how $\left \langle (\delta \tilde{z})^2 \right \rangle$ evolves with time. Since we focus on the diffusion behavior of particles on small scales, we trace the particles before $\left \langle (\delta \tilde{z})^2 \right \rangle$ reaches $L$.

\begin{figure*}[htbp]   
    \subfigure[]{
     \includegraphics[width=8cm]{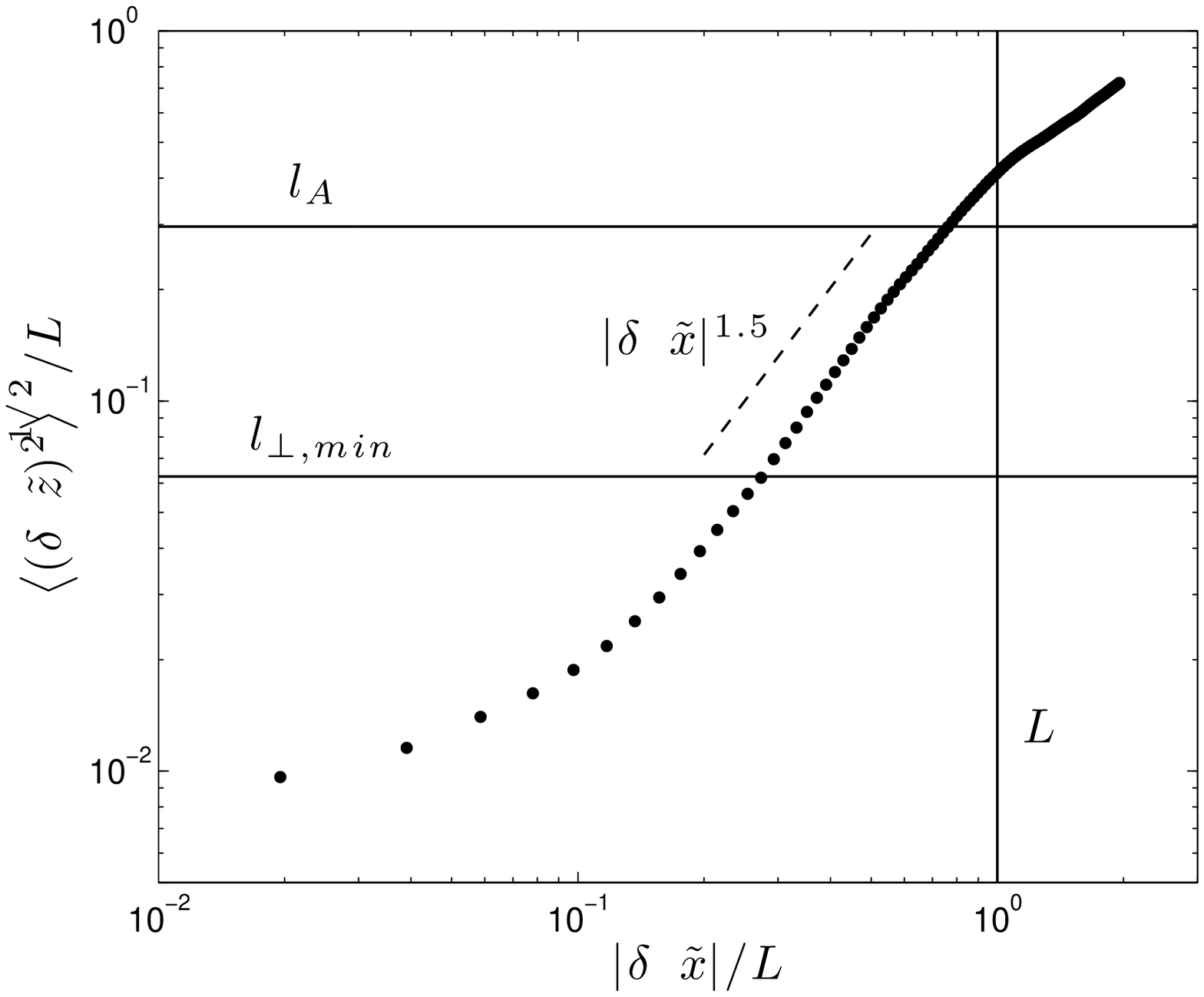}\label{fig8a}}
    \subfigure[]{
     \includegraphics[width=8cm]{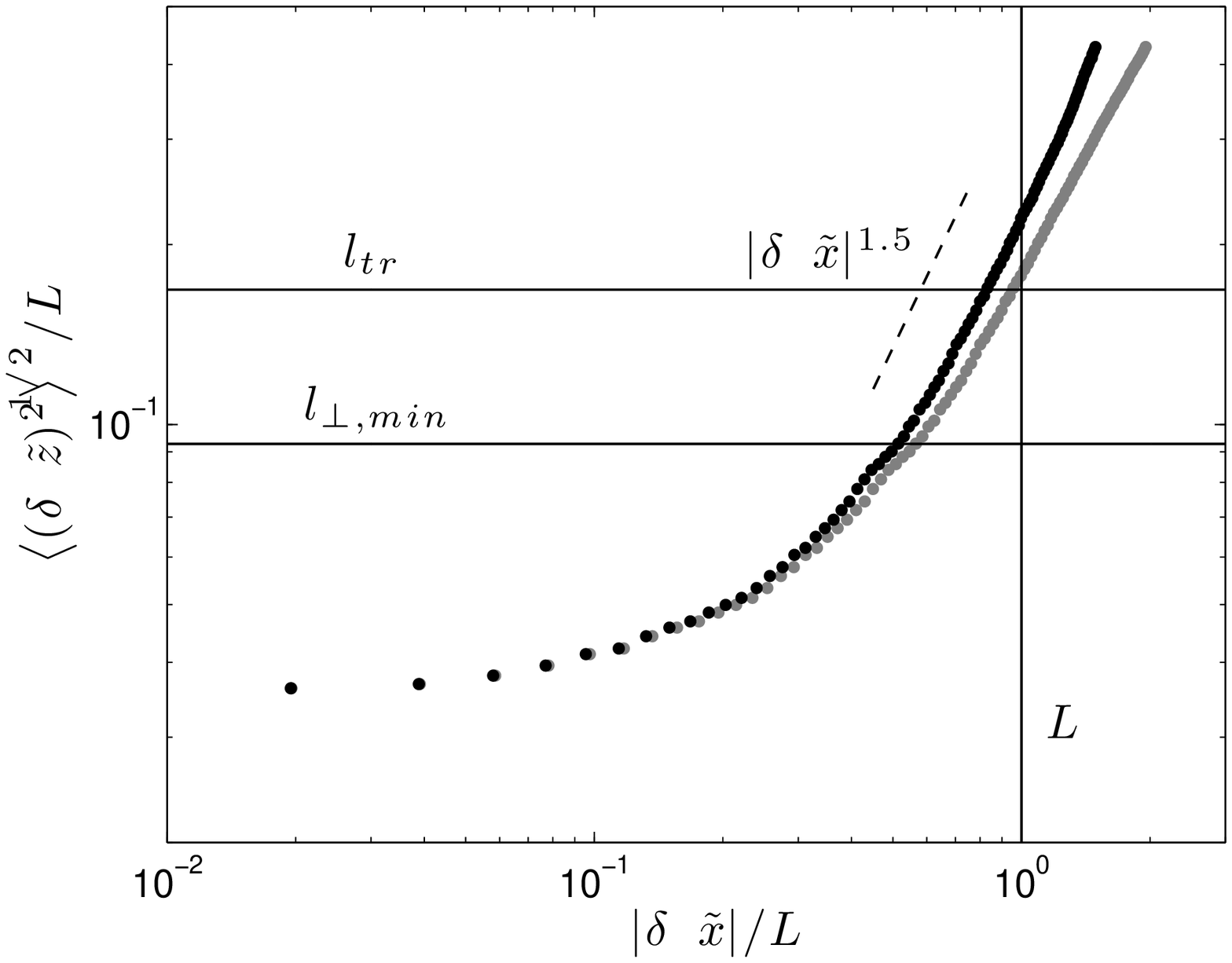}\label{fig8b}}
   \caption{\small  $\left \langle (\delta \tilde{z})^2 \right \rangle^{1/2}/L$ vs. $|\delta \tilde{x}|/L$ for (a) $M_A=1.5$ and (b) $M_A=0.41$. (b) shows the plots using constant $u$ (lower profile) and $u_\parallel$ (upper profile). The dashed line indicates the slope of the curve.}
\end{figure*}

Since the particles in the sub-Alfv$\acute{\rm e}$nic turbulence usually have $\lambda_\parallel$ larger than the injection scale in our simulations (see Section 5), we assume that particles move strictly along magnetic field lines during the simulation. We then determine the distance travelled along magnetic field lines by 
\begin{equation}
         \delta \tilde{x}=u \delta t,
\label{vel}         
\end{equation}
where $u$ is a constant velocity derived from the initial Larmor radius, and $\delta t$ is the corresponding time. Fig.~\ref{fig8a} and \ref{fig8b} display $\left \langle (\delta \tilde{z})^2 \right \rangle^{1/2}/L$ as a function of the displacement of particles moving along the field $|\delta \tilde{x}|/L$ for super-Alfv$\acute{\rm e}$nic and sub-Alfv$\acute{\rm e}$nic cases. The separation grows as distance along the field lines to the $1.5$ power after passing the minimum perpendicular scale of eddies $l_{\perp, min}$, up to the injection scale of the strong MHD turbulence ($l_A=L/M_A^3$ for $M_A>1$ and $l_{\rm tr}\sim LM_A^2$ for $M_A<1$, 
\citealt{Lazarian06}; YL08). 
Our result is also consistent with earlier studies on the separation of field lines in 
\citet{LVC04} and \citet{Maron04}. This consistency verifies that particle superdiffusion on small scales is determined by the divergence of field lines related to Richardson spatial diffusion \citep[see][]{LY13}. 

Notably, in real astrophysical world, since CRs have $r_L$ much larger than $l_{\perp, min}$, they always exhibit superdiffusion on scales smaller than $L$. 

Besides the general relation between  $\left \langle (\delta \tilde{z})^2 \right \rangle^{1/2}$ and $|\delta \tilde{x}|$ we confirmed, we also consider a specific case with $\lambda_\parallel<L$. To study the regime $\lambda_\parallel<L$, applicable to most of the CRs, we add resonant slab fluctuations to the initial turbulent magnetic fields obtained through MHD turbulence simulations. Since the slab component is very efficient in pitch-angle scattering through gyroresonance, $\lambda_\parallel$ can be effectively reduced to values smaller than the injection scale of the turbulence with sufficient amplitude. 
This addition will not affect the statistical properties of particle transport across the field for the following reasons. First of all, the small scale resonant slab modes are uncorrelated with the original turbulence modes. Moreover, the contribution of slab modes to particle cross field transport is sub-diffusive 
\citep[see][]{Kota_Jok2000} 
and therefore can be neglected.

\begin{figure}[htbp] 
   \centering 
      \includegraphics[width=8cm]{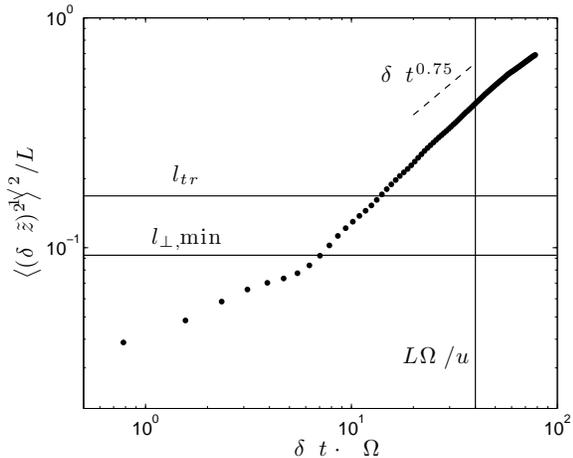}
   \caption{\small  The same as Fig.~\ref{figev} but for $\lambda_\parallel<L$. The dashed line indicates the slope of the curve.}
   \label{fig9}
\end{figure}

Fig.~\ref{fig9} displays $\left \langle (\delta \tilde{z})^2 \right \rangle^{1/2}$ as a function of time in this case. For the scales $L>|\delta \tilde{x}|>\lambda_\parallel$, our result suggests
\begin{equation}
         \left \langle (\delta \tilde{z})^2 \right \rangle^{1/2} \propto t^{0.75},
\end{equation}
consistent with YL08 predictions (see equation (30) and (31) in YL08).
\\

\begin{figure}[htbp] 
   \centering 
      \includegraphics[width=8cm]{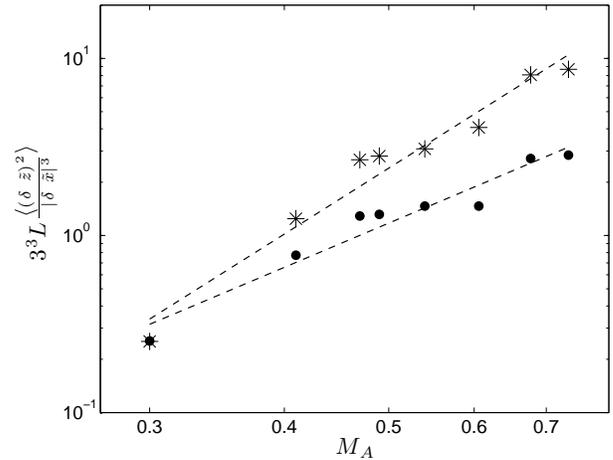}
   \caption{\small  $\frac{\left \langle (\delta \tilde{z})^2 \right \rangle}{|\delta \tilde{x}|^3}$ as a function of $M_A$. Filled circles are for previous results with constant velocity, and asterisks are for results after the correction for $u$. The dashed lines are the best fits to the data. }
   \label{fig10}
\end{figure}

Fig.~\ref{fig10} presents the ratio $\frac{\left \langle (\delta \tilde{z})^2 \right \rangle}{|\delta \tilde{x}|^3}$ as a function of $M_A$. The best fit to the numerical data shows 
\begin{equation}
         \frac{\left \langle (\delta \tilde{z})^2 \right \rangle}{|\delta \tilde{x}|^3}\propto M_A^{2.58\pm0.64}
\end{equation}
at 95\% confidence. Actually, we notice that even for particles with $\lambda_\parallel \geqslant L$, the pitch angles change significantly especially in cases with higher $M_A$ values. The assumption of no pitch-angle scattering may lead to an overestimate of the parallel distances. Thus, we replace $u$ with the parallel particle velocity $u_\parallel$ in Eq.~(\ref{vel}).  We derive the following relation after this correction (also see Fig.~\ref{fig8b} and Fig.~\ref{fig10}), 
\begin{equation}
         \frac{\left \langle (\delta \tilde{z})^2 \right \rangle}{|\delta \tilde{x}|^3}\propto M_A^{3.84\pm0.78}
\end{equation}
with 95\% confidence, in good agreement with the theoretical predictions
(\citealt{LV99}; YL08).

\section{Discussion}
The spatially dependent CR diffusion we obtain with physically motivated model of turbulence should help resolve the current observational puzzles in relation to CR propagation in various astrophysical environments. Our conclusions on CR diffusion will contribute to a fundamental understanding of the underlying processes of non-local observables, like CR anisotropy and galactic $\gamma$-ray diffuse emission.

Different from simulations on CR transport that employ a \emph{slab/2D composite model} and synthetic turbulence data
(e.g., \citealt{Giacalone_Jok1999, Qin2002, Tau13}), 
we use direct 3D MHD numerical simulation to produce turbulence data cube. The reason are as follows. First, \emph{slab/2D composite} approximation of turbulence is not supported by numerical simulations. Also, it has not been realized to generate the numerically confirmed scale-dependent anisotropy with respect to the local magnetic field in synthetic turbulence
\citep{LV99, CV00}. 
Performing test particle simulations in the turbulence that misses essential physics does not lead to reliable results on CR diffusion.

Perpendicular diffusion of CRs across the mean magnetic field has been considered a difficult problem of particle astrophysics for a long time. We for the first time demonstrated numerically that perpendicular transport of CRs depends on $M_A^4$ of the turbulence. Our numerical results can be used for a wide range of applications. On large scales, our results on perpendicular diffusion can be applied to depict the normal diffusion of energetic particles in heliosphere, with strong observational constraints
\citep{Maclennan2001}. 
 On scales smaller than the energy injection scale, the superdiffusive process is important for describing propagation and acceleration of CRs in supernovae shells and shock regions.
 \citet{LY13} find that the superdiffusive behavior of CRs can change the properties of CR acceleration in shocks, and decrease efficiency of CR acceleration in perpendicular shocks. Our numerical confirmation of the superdiffusive behavior provides an additional justification for the theory above. The superdiffusive transport on small scales we obtained can also naturally explain the experimental data in heliosphere. For instance, superdiffusion of solar energetic particles has been argued based on the analysis of the particle time profiles 
\citep{Perri2009}. 
They find the propagation of energetic particles in the interplanetary space is superdiffusive. 

The feedback of CRs on turbulence, such as gyroresonance instability 
\citep[see][]{YL11} is not included in the test particle simulations. This shall be a subject of future study.

The diffusion processes we studied in this work have important implications for other issues. Similar diffusion properties and the dependence on $M_A$ can also be applied to thermal particles and our numerical results are consistent with the analytical descriptions in 
\citet{Lazarian06}.
The thermal diffusion has a profound impact on problems such as cooling flows in clusters of galaxies. 

As a fundamental astrophysical process, magnetic reconnection is controlled by the turbulent wandering of magnetic field 
\citep{LV99}. 
The diffusion behavior of field lines is essential for determining the reconnection rate in turbulent medium.  The $M_A$ dependence that we for the first time numerically confirmed in this paper can help quantitively determine the extension degree of the outflow region and the resulting magnetic reconnection rate. 

\section{summary}
We provide a realistic description of particle transport with test particle simulations in tested model of compressible turbulence. Our results are in general consistent with the nonlinear transport theory developed in YL08 and can be summarized as below :
\begin{enumerate}
\item Pitch-angle scattering experiments are consistent with nonlinear theory, showing the dominance of mirror interaction for most of the pitch angles except for small ones. 
\item The nonlinear effect for pitch angles close to 90$^\circ$ has been confirmed by our simulations.
\item We have demonstrated numerically that CRs are diffusive on large scales. We show that perpendicular diffusion coefficient depends on $M_A^4$ in the case of $\lambda_\| >L$ in sub-Alfv\'enic turbulence.
\item On small scales, CRs experience superdiffusion.
\end{enumerate}

\begin{acknowledgements}
SX and HY are supported by NSFC grant AST-11073004. We acknowledge the computing support from FSC-PKU. We have benefited from valuable discussions with Blakesley Burkhart,  Alex Lazarian and Shangfei Liu. We also thank the anonymous referee for their helpful suggestions.
\end{acknowledgements}

\bibliographystyle{apj.bst}
\bibliography{yan}
\end{document}